\documentclass[twocolumn,notitlepage,prb,superscriptaddress,longbibliography,nofootinbib]{revtex4-2}
\usepackage{amsfonts}
\DeclareUnicodeCharacter{2032}{\ensuremath{\prime}}
\DeclareUnicodeCharacter{2212}{\ensuremath{-}}
\DeclareUnicodeCharacter{03B4}{\ensuremath{\delta}}
\DeclareUnicodeCharacter{22A5}{\ensuremath{\perp}}
\DeclareUnicodeCharacter{2225}{\ensuremath{\parallel}}
\usepackage{textcomp}
\usepackage{times}
\usepackage{graphicx}
\usepackage{float}
\usepackage{latexsym,amsmath,amssymb,bm,euscript}
\usepackage[dvipsnames]{xcolor}
\usepackage{subfigure}
\usepackage{epstopdf}
\usepackage[colorlinks=true,linkcolor=blue,citecolor=blue,urlcolor=blue]{hyperref}
\usepackage{hyperref}
\usepackage{soul}
\usepackage[normalem]{ulem}
\usepackage{mathrsfs}
\usepackage{amsmath}
\usepackage{amstext}
\usepackage[bottom]{footmisc}
\usepackage{amsbsy}
\usepackage{ulem}
\usepackage{booktabs}

\usepackage{xcolor}
\usepackage[urlcolor=blue]{hyperref}      
\hypersetup{
    colorlinks = true,                    
    citecolor = {blue},
    linkcolor = {purple},
           }
\newcommand{\be}{\begin{equation}}
\newcommand{\ee}{\end{equation}}
\newcommand{\bea}{\begin{eqnarray}}
\newcommand{\eea}{\end{eqnarray}}

\begin{document}	
\title{Evolution from intralayer to interlayer superconductivity in a bilayer $t$-$J$-$J_{\bot}$ ladder}

\author{Yuan Yang}
\affiliation{Great Bay Institute for Advanced Study, Dongguan 523000, China}
\affiliation{School of Physical Sciences, Great Bay University, Dongguan 523000, China}
\affiliation{Department of Modern Physics, University of Science and Technology of China, Hefei, Anhui 230026, China}

\author{Xin Lu}
\affiliation{School of Physics, Beihang University, Beijing 100191, China}

\author{Yuan Wan}
\affiliation{Institute of Physics, Chinese Academy of Sciences, Beijing 100190, China}

\author{Wei-Qiang Chen} 
\email{chenwq@sustech.edu.cn} 
\affiliation{State Key Laboratory of Quantum Functional Materials, Department of Physics, and Guangdong Basic Research Center of Excellence for Quantum Science, Southern University of Science and Technology, Shenzhen 518055, China}
\affiliation{Quantum Science Center of Guangdong-Hong Kong-Macao Greater Bay Area, Shenzhen 518045, China}

\author{Shou-Shu Gong}
\email{shoushu.gong@gbu.edu.cn}
\affiliation{School of Physical Sciences, Great Bay University, Dongguan 523000, China}
\affiliation{Great Bay Institute for Advanced Study, Dongguan 523000, China}

\date{\today}
\begin{abstract}
Motivated by the bilayer cuprate superconductors and nickelate superconductor La$_3$Ni$_2$O$_7$, we investigate the evolution from intralayer to interlayer superconductivity in finite-width bilayer $t$-$J$-$J_{\bot}$ ladders, where $t$ is the in-plane electron hopping, $J$ is the in-plane spin  interaction, and $J_{\bot}$ is the inter-plane spin interaction. 
By means of the density matrix renormalization group calculations, we map out the quantum phase diagram of the bilayer two-leg ladders by tuning $J_{\bot}$ in a large doping range $\delta = 1/8 - 1/2$.
We find that a large $J_{\bot}$ can always drive an interlayer superconductivity by coupling the two layers in both the Luther-Emery liquid and Luttinger liquid states.
By coupling two Luther-Emery liquid states, the in-plane superconductivity evolves to inter-plane superconductivity either through an intermediate charge density wave (CDW) phase or directly, depending on doping ratio.
This emergent CDW phase, which exists over a finite doping range, appears to develop from the CDW state of the two-leg ladder at $\delta = 1/4$.
By coupling two Luttinger liquids, the in-plane Luttinger liquids show a transition to the inter-plane superconducting phase at large $J_{\bot}$, as reported in previous literature. 
Interestingly, in the intermediate $J_{\bot}$ regime we find that while the in-plane Luttinger-liquid features remain stable, the inter-plane superconductivity can develop an enhanced quasi-long-range order with the power exponent $K^{zz}_{\rm SC} \sim 1$.
At last, we show that the interlayer superconductivity is also stable by coupling the bilayer three-leg $t$-$J$ ladders by a strong $J_{\bot}$ interaction, from both the Luther-Emery liquid and Luttinger-liquid states.
\end{abstract}
\maketitle

\section{Introduction}
Since the discovery of the cuprate superconductors, understanding the exotic phases in the cuprates phase diagram has been one of the central topics in condensed matter physics~\cite{keimer2015quantum,lee2006doping,scalapino2012common,armitage2010progress}.  
For the single-layer cuprate compounds such as La$_2$CuO$_4$~\cite{bednorz1986possible,czyzyk1994local} and HgBa$_2$CuO$_4$~\cite{putilin1993superconductivity}, the essential low-energy physics may be captured by the single-band square-lattice Hubbard model and the related $t$-$J$ model with the no-double-occupancy constraint~\cite{hubbard1963electron,izyumov1997strongly,zhang1988effective}.
These microscopic models have been studied intensively to investigate the novel phases related to the cuprates, such as the $d$-wave superconductivity (SC), pseudogap, and non-Fermi-liquid behaviors~\cite{lu2023emergent,lu_sign_structure,tJ_Feng_2023,jiang2021ground,jiang2021high,gong2021robust,chung2020plaquette}.

Experimentalists have also found bilayer cuprate superconductors, such as YBa$_2$Cu$_3$O$_{6.6}$~\cite{wu1987superconductivity,tranquada1992neutron,rossat1991neutron,mook1993polarized}, La$_2$CaCu$_2$O$_6$~\cite{cava1990superconductivity},
Pb$_2$Sr$_2$YCu$_3$O$_8$~\cite{cava1988superconductivity}, and EuSr$_2$Cu$_2$NbO$_8$~\cite{babu1991electrical,fan2024superconductivity,sakakibara2014orbital}.
These compounds may exhibit enhanced superconducting transition temperature and modified spectral properties, suggesting an important role of the interlayer coupling~\cite{fan2024superconductivity,sakakibara2014orbital,goto1998roles,zaleski2005dependence,ubbens1994spin,medhi2009phase}. 
Theoretically, these compounds are often modeled by extending the $t$-$J$ model to include interlayer electron hopping $t_{\bot}$ and spin exchange interaction $J_{\bot}$.
Based on these models, many unusual properties of the bilayer cuprates have been explored, such as the spin-gap phase, the coexistence of antiferromagnetic order and $d$-wave SC, and the possible planar flux phase~\cite{ubbens1994spin,medhi2009phase,yamase2011multilayer,liechtenstein1995s,kuboki1995energy,eder1995ground,zhao2005t,gimm1997bilayer,yin2000superconductivity}.
Both exact diagonalization and mean-field calculations found that a large $J_{\bot}$ is helpful for the inter-plane pairing~\cite{ubbens1994spin,eder1995ground}, and the in-plane $d$-wave SC may coexist with the inter-plane $s$-wave SC in some parameter regime~\cite{ubbens1994spin}.

Recently, a significant breakthrough in the study of unconventional SC is the discovery of the nickel-based superconductors, including the infinite-layer nickelates such as Nd$_{1-x}$Sr$_x$NiO$_2$ ~\cite{li2019superconductivity,chow2025bulk}, the bilayer nickelates such as La$_3$Ni$_2$O$_7$ ~\cite{sun2023signatures,zhang2024high,wang2024bulk,wang2024pressure,XIE20243221,li2024pressure,yang2024orbital,dong2024visualization,chen2024electronic,wang2024normal,chen2024evidence,wang2024normal,Hepting} and the trilayer nickelates such as La$_4$Ni$_3$O$_{10}$ ~\cite{zhu2024superconductivity,zhang2025superconductivity}. 
Among them, La$_3$Ni$_2$O$_7$ shows a SC below the temperature $T_{c\text{-}{\rm onset}} \simeq 80 K$ and features two NiO$_2$ layers, which has some important similarities to the bilayer cuprates including the splitting of the main band induced by the interlayer interaction and the Fermi surface reconstruction under pressure~\cite{sakakibara2014orbital,fan2024superconductivity}.
Density functional theory studies have provided crucial insights into the electronic structure of La$_3$Ni$_2$O$_7$ under pressure~\cite{luo2023bilayer}, which find that the $d_{z^2}$ orbitals contribute to the $\gamma$ hole pocket near the Fermi level and the interlayer coupling between the $d_{z^2}$ orbitals may play an crucial role in driving the SC in the pressurized phase~\cite{luo2023bilayer,zhang2023electronic,lechermann2023electronic,gu2025effective,yang2023possible,christiansson2023correlated,lechermann2024electronic,jiang2024high,liao2023electron,luo2024high,wu2024superexchange,shen2023effective,QiangHua2025}. 
This important role of interlayer coupling for SC has also been unveiled in model studies~\cite{shen2023effective,gu2025effective,sakakibara2024possible,lechermann2023electronic,zhang2023electronic,yang2023possible,luo2023bilayer,qu2024bilayer,lu2024interlayer,christiansson2023correlated,cao2024flat,wu2024superexchange,luo2024high,liao2023electron,liu2023s,oh2023type,qin2023high,pan2024effect,jiang2024high,huang2023impurity,kaneko2024pair,lange2024feshbach,lu2023superconductivity,lechermann2024electronic,zhang2024strong,chen2024orbital,qu2023roles,jiang2024pressure,schlomer2024superconductivity,fan2024superconductivity,wang2024normal,QiangHua2025,duan2025orbital,liao2024orbital,braz2025interlayer,ma2022doping,capponi2004current}.
While many different models have been proposed to study the SC of La$_3$Ni$_2$O$_7$~\cite{shen2023effective,gu2025effective,bejas2025out,liao2023electron,liao2024orbital,qu2024bilayer,duan2025orbital,lu2024interlayer,liu2023s,cao2024flat,qin2023high,oh2023type,kaneko2024pair,qu2023roles,fan2024superconductivity}, a theory pointed out the strong Hund's coupling between $d_{x^2-y^2}$ and $d_{z^2}$ orbitals~\cite{lu2024interlayer,luo2024high,liu2023s,liu2023s,qin2023high,tian2024correlation} and obtaind a bilayer single-band $t$-$J$-$J_{\bot}$ model by integrating out the nearly half-filled $d_{z^2}$ orbitals in the strong Hund’s coupling and large-$U$ limits~\cite{lu2024interlayer,qu2024bilayer,fan2024superconductivity,qu2023roles,pan2024effect}.
Recent studies on the bilayer $t$-$J$-$J_{\bot}$ model have been focused on the doping ratio $\delta \sim 0.5$, which is close to the electron filling of La$_3$Ni$_2$O$_7$~\cite{qu2024bilayer}. 
Mean-field calculations have demonstrated a $J_{\bot}$-driven transition from the intralayer $d$-wave pairing to interlayer $s$-wave pairing~\cite{lu2024interlayer}, which may also include a narrow intermediate $s$+$id$-wave state.
The large-$N$ calculation beyond mean-field level also finds a possible out-of-plane bond-order phase triggered by $J_{\bot}$~\cite{bejas2025out}, which may compete with the inter-plane SC. 
Numerically, it has been identified that a large $J_{\bot}$ can stabilize an inter-plane SC in the bilayer two- and three-leg ladders~\cite{qu2023roles,chen2024orbital,zhang2024strong,kaneko2024pair,oh2023type,qu2024bilayer,sakakibara2024possible}. 
With these progress in the understanding of the bilayer $t$-$J$ model, there are still many interesting questions that are not very clear.
For example, how does the in-plane SC evolve to the inter-plane SC with growing $J_{\bot}$?
Whether the in-plane $d$-wave SC can coexist with inter-plane SC? 
Addressing these questions may be helpful for better understanding of bilayer cuprate and nickelate superconductors.

\begin{figure*}[htb]
\includegraphics[width=0.7\linewidth]{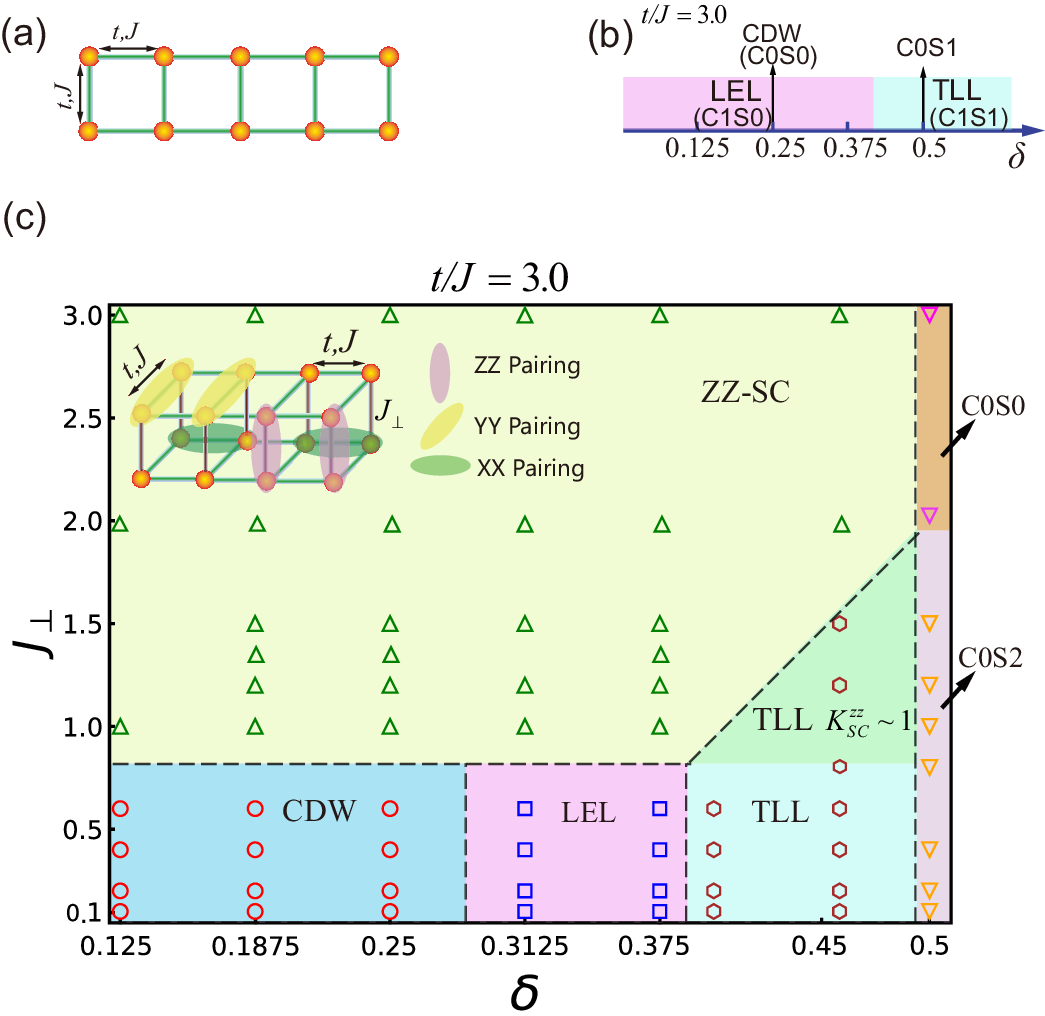}
\caption{Model Hamiltonian and quantum phase diagram.
(a) Schematic figure of a two-leg $t$-$J$ ladder. (b) Phase diagram of the two-leg $t$-$J$ model ($t/J = 3$) as a function of doping ratio $\delta$~\cite{lu2023ground}. With increased $\delta$, the system has the Luther-Emery liquid (LEL) phase and Tomonaga-Luttinger liquid (TLL) phase. At the special doping levels $\delta = 1/4$ and $1/2$, the system is the charge density wave (CDW) and the C0S1 states, respectively.
Here, ``C'' and ``S'' denote charge and spin sectors, respectively.
C0S1 means there is a gapless mode in the spin (S) sector and no gapless mode in the charge (C) sector.
(c) $J_\bot-\delta$ phase diagram obtained for the bilayer two-leg $t$-$J$-$J_{\bot}$ ladders with a fixed ratio of $t/J=3$ ($J_{\bot}/J \geq 0.1$, $\delta = 1/8 - 1/2$). 
The inset illustrates the lattice structure of the bilayer two-leg $t$-$J$-$J_{\bot}$ model, showing the three types of spin-singlet pairings along the different directions: XX, YY and ZZ pairings. At large $J_{\bot}$, the system enters to the inter-plane SC phase, denoted as ZZ-SC and marked by the triangle symbol. The circles mark the CDW phase, the rhombuses denote the LEL phase, and the hexagons denote the TLL phase. At $\delta = 1/2$, the system is the C0S2 state at weak $J_{\bot}$ (two copies of the C0S1 state in subfigure (b)) and a fully gapped C0S0 state at strong $J_{\bot}$ [see Appendix~\ref{subsec::C0S1}]. For the TLL phase, the inter-plane SC is enhanced at the intermediate $J_{\bot}$ (with the power exponent $K^{zz}_{SC} \sim 1$), before the phase transition to the ZZ-SC phase. Notice that the symbols represent the commensurate filling points that we have simulated. We do not accurately determine the phase boundaries but use the dashed lines to denote the rough boundaries.}
\label{fig::lattice}
\end{figure*}

\begin{figure*}[htb]
\includegraphics[width=1.0 \linewidth]{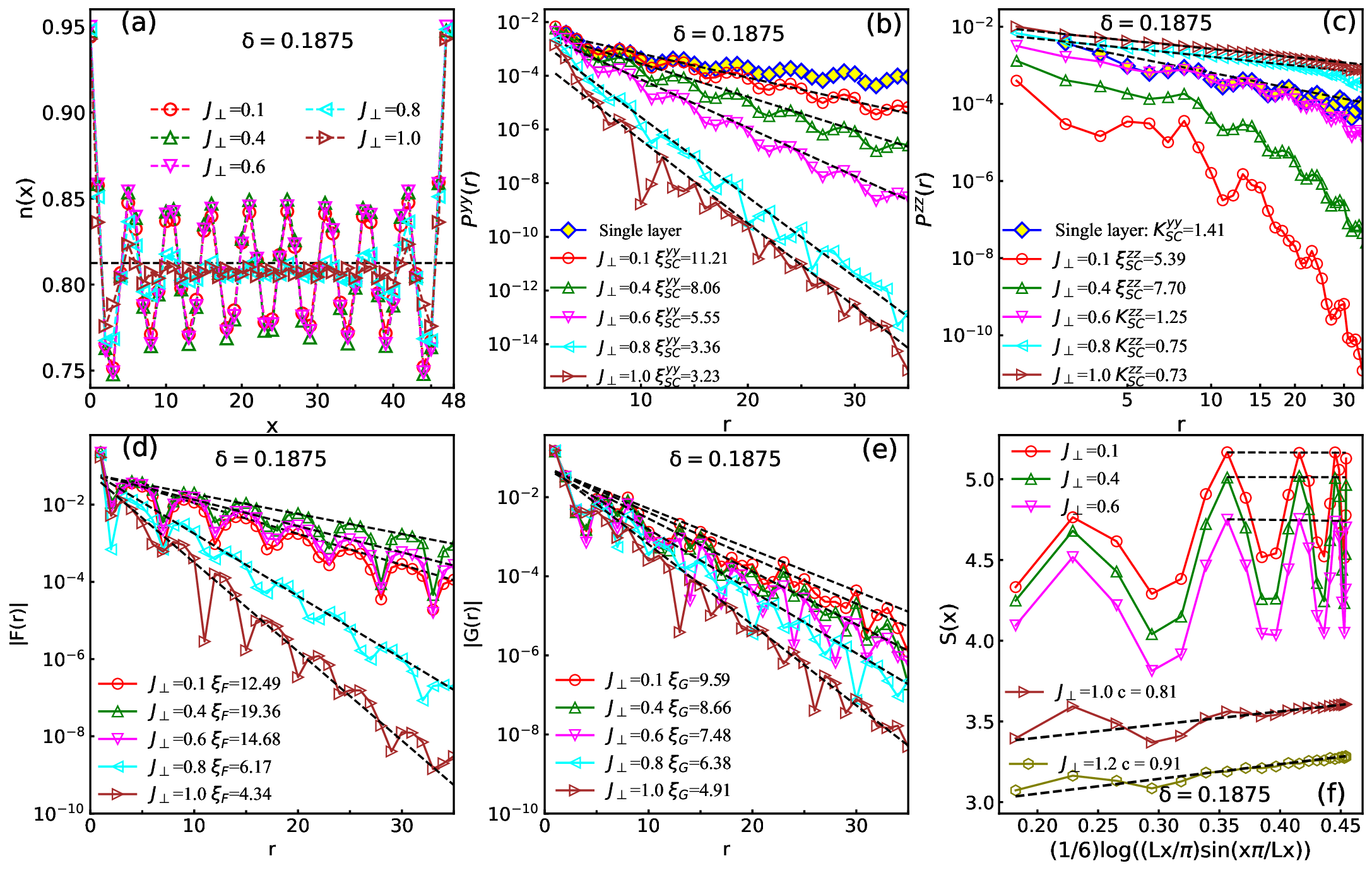}
\caption{Measurements of the bilayer two-leg $t$-$J$-$J_{\bot}$ model at $\delta=0.1875$. (a) Charge density profile $n(x)$. (b) In-plane pairing correlation $P^{yy}(r)$ shown in the semi-logarithmic scale. (c) Inter-plane pairing correlation $P^{zz}(r)$ shown in the double-logarithmic scale. (d) Spin correlation $F(r)$ shown in the semi-logarithmic scale. (e) Single-particle Green's function $G(r)$ shown in the semi-logarithmic scale. (f) Entanglement entropy $S(x)$ versus the conformal distance. The central charge $c$ is extracted using Eq.~\eqref{eq::Sx}. 
To reduce the open boundary effects, the bulk entropy is selected for the fitting. In the CDW phase, we choose to fit the upper envelope of the bulk entropy, as shown by the dashed lines at $J_{\perp} / J = 0.1, 0.4, 0.6$, which indicate the nearly flat envelope that agrees with a fully gapped CDW state.
The system length is $L_x=48$, and the number of kept SU(2) multiplets is set between $6000$ and $12000$.
}
\label{fig::01875_1}
\end{figure*}

\begin{figure*}[htb]
\includegraphics[width=1.0 \linewidth]{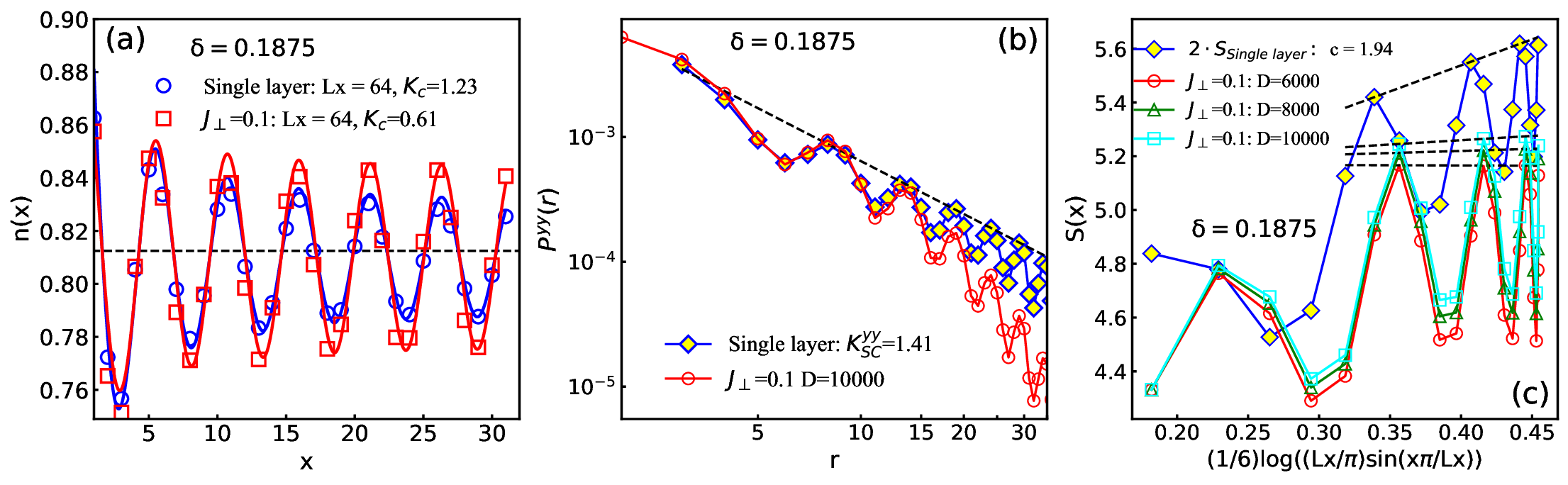}
\caption{Comparison of (a) local charge density $n(x)$, (b) in-plane pairing correlation $P^{yy}(r)$, and (c) entanglement entropy $S(x)$ between the two-leg $t$-$J$ model and the bilayer $t$-$J$-$J_{\bot}$ model at $\delta=0.1875$. In (c), the blue line with yellow squares denote twice of the entanglement entropy of the single-layer model, serving as a reference for the weakly coupled bilayer system at $J_{\bot}/J=0.1$. Other curves show $S(x)$ for the bilayer model at different SU(2) multiplet bond dimensions. The central charge $c$ is extracted using Eq.~\eqref{eq::Sx}. 
}
\label{fig::01875_2}
\end{figure*}

To further explore these questions, we systematically study the ground-state properties of finite-width bilayer $t$-$J$-$J_{\bot}$ ladders using density matrix renormalization group (DMRG) calculations.
Based on the established quantum phases of the two-leg $t$-$J$ ladder [see Figs.~\ref{fig::lattice}(a) and \ref{fig::lattice}(b)]~\cite{lu2023ground},we obtain a rich quantum phase diagram of the bilayer two-leg $t$-$J$ ladders coupled by $J_{\bot}$. 
By tuning a large doping range $\delta = 0.125 - 0.5$ and $J_\bot$, we identify several phases as shown in Fig.~\ref{fig::lattice}(c). 
In the presence of a relatively large $J_{\bot}$, the inter-plane SC always emerges except at the special doping level $\delta = 0.5$, owing to the combined effect of filling factor and lattice geometry~\cite{white2002friedel,lu2023ground}. 
At smaller doping range, where the two-leg ladder is the Luther-Emery liquid, the in-plane SC evolves to the inter-plane SC either through intermediate charge density wave (CDW) states or directly, depending on doping ratio.
These CDW states appear to develop from the CDW state of the two-leg ladder at $\delta = 1/4$, which seems to expand to more doping levels in this bilayer system.
When the in-plane SC evolves to inter-plane SC directly, our DMRG data find no evidence of a coexistence.
At larger doping range when the two-leg ladder is the Luttinger liquid, a strong inter-plane SC develops when the Luttinger liquids are gapped by a relatively large $J_{\bot}$.
At the intermediate $J_{\bot}$ when the in-plane Luttinger-liquid features remain stable, we find a quasi-long-range inter-plane SC with the power exponent of the interlayer pairing correlation function $K^{zz}_{\rm SC} \sim 1$, showing the greatly enhanced inter-plane SC in the Luttinger liquid phase.
For the wider system size by coupling the bilayer three-leg $t$-$J$ ladders, we find that the inter-plane SC observed near $\delta = 0.5$ (including $\delta = 0.5$) at large $J_{\bot}$~\cite{qu2024bilayer} is also stable at the smaller doping range, suggesting that the inter-plane SC phase can be stabilized by strong $J_{\bot}$ over a large doping range.

The paper is organized as follow.
In Sec.~\ref{sec::model}, we introduce the model Hamiltonian and the details of DMRG calculations.
In Sec.~\ref{sec::result}, we demonstrate the DMRG results that identify the different quantum phases.
The last Sec.~\ref{sec::summary} is devoted to the summary and discussion.

\section{Model and Method}
\label{sec::model}

The Hamiltonian of the bilayer $t$-$J$-$J_{\bot}$ model is defined as:
\begin{eqnarray}
    H &=& -t\sum_{\langle i,j\rangle,\mu,\sigma} (c^{\dagger}_{i,\mu,\sigma} c_{j,\mu,\sigma} + \rm{H.c.}) \nonumber \\ 
    &\;& + J\sum_{\langle i,j \rangle, \mu} ({\bf S}_{i,\mu} \cdot {\bf S}_{j,\mu} - \frac{1}{4}n_{i,\mu}n_{j,\mu}) \nonumber \\
    &\;& + J_\bot\sum_{i}{\bf S}_{i,1} \cdot {\bf S}_{i,2}, 
\label{eq::Hamiltonian}
\end{eqnarray}
where $c^{\dagger}_{i,\mu,\sigma}$ ($c_{i,\mu,\sigma}$) is the spin-$\sigma$ ($\sigma=\pm1/2$) electron creation (annihilation) operator at the site $i$ in the layer $\mu$ ($\mu = 1,2$), ${\bf S}_{i,\mu}$ is the spin-$1/2$ operator, and ${n}_{i,\mu} = \sum_{\sigma} {c}^{\dagger}_{i,\mu,\sigma} {c}_{i,\mu,\sigma}$ is the electron number operator.
For each site, the no-double-occupancy constraint is implemented.
$t$ is the intralayer nearest-neighbor hopping integral.
$J$ and $J_{\bot}$ are the intralayer and interlayer spin exchange couplings, respectively. 
In our study, we choose $t/J = 3$ to mimic a large Hubbard $U$, and we tune $J_{\bot} / J$ and doping ratio $\delta = 0.125 - 0.5$.
Given the challenges of numerical simulations at very weak $J_{\bot}$, we consider $J_
\bot / J \geq 0.1$.
We mainly study the bilayer two-leg ladders with $L_y = 2$ in each layer, and also consider the wider systems with three-leg ($L_y = 3$) ladders in each layer at large $J_{\bot}$.
We use $L_x$ to denote the length of the system, and we have studied different lengths from $L_x = 48$ to $L_x = 64$, which can help us to reduce finite-size effect.

We solve the ground state of the model by using the density matrix renormalization group (DMRG) calculations~\cite{white1992density}.
We implement the spin-rotational SU(2)~\cite{mcculloch2007density} and charge U(1) symmetries. 
We keep the bond dimensions $6000 - 8000$ for most calculations, and up to $D = 14000$ SU(2) multiplets for some jobs that need high accuracy.
The DMRG truncation error is about $10^{-6}$, which gives accurate measurements.

To characterize the different quantum states, we compute the charge density profile and various correlation functions.
We use the averaged charge density for each column $n(x)$ to characterize the charge density order, which is defined as:
\begin{equation}
n(x) = \frac{1}{2 L_y}\sum_{\mu=1}^{2}\sum_{y=1}^{L_y}\sum_{\sigma}\langle c_{(x,y),\mu,\sigma}^{\dagger} c_{(x,y),\mu,\sigma}\rangle,
    \label{eq::nx} 
\end{equation}
where $(x,y)$ denotes the position of the site in each layer.
The spin-spin correlation function is defined as:
\begin{equation}
F(r) = \frac{1}{2L_y}\sum_{\mu=1}^{2}\sum_{y=1}^{L_y}\langle {\bf S}_{(x_0,y),\mu} \cdot {\bf S}_{(x_0+r,y),\mu} \rangle,
\label{eq::Fr}
\end{equation}
where $(x_0,y),\mu$ is the reference site $(x_0,y)$ in the layer $\mu$. 
The single-particle Green's function is defined as:
\begin{equation}
G(r) = \frac{1}{2L_y}\sum_{\sigma}\sum_{\mu=1}^{2}\sum_{y=1}^{L_y}\langle c_{(x_0,y),\mu,\sigma}^{\dagger}c_{(x_0+r,y),\mu,\sigma} \rangle
\label{eq::Gr},
\end{equation}
and the density-density correlation function is expressed as:
\begin{eqnarray}
D(r) &=& \frac{1}{2L_y}\sum_{\mu=1}^{2}\sum_{y=1}^{L_y} \big ( \langle n_{(x_0,y),\mu} n_{(x_0+r,y),\mu} \rangle - \nonumber \\ 
&\;& \langle n_{(x_0,y),\mu}\rangle\langle n_{(x_0+r,y),\mu} \rangle \big ).
\label{eq::Dr}
\end{eqnarray}
For characterizing SC, we calculate the spin-singlet pairing correlation function
\begin{equation}
P^{\alpha\alpha}(r) = \langle \Delta_{\alpha}^{\dagger}(x_0,y,z) \Delta_{\alpha}(x_0+r,y,z) \rangle,
\label{eq::Pr}
\end{equation}
where $\alpha\alpha = xx,yy,zz$ denote the bond orientations as shown in Fig.~\ref{fig::lattice}(c).
$\Delta^{\dagger}_{\alpha} = (c^{\dagger}_{(x,y,z)\uparrow}c^{\dagger}_{(x,y,z)+e_{\alpha}\downarrow} - c^{\dagger}_{(x,y,z)\downarrow}c^{\dagger}_{(x,y,z)+e_{\alpha}\uparrow} )/\sqrt{2}$
is the spin-singlet pair-field creation operator, where $e_{\alpha}(\alpha=x,y,z)$ denotes the unit length along the $x$, $y$ and $z$ directions. 
To diminish the boundary effect, we measure the correlation functions by choosing the reference site at the position $x_0 \sim L_x/4$.

\section{DMRG results}
\label{sec::result}

We first briefly review the phase diagram of the two-leg $t$-$J$ ladder at $t/J = 3$~\cite{chen2018two,lu2023ground}, as shown in Figs.~\ref{fig::lattice}(a) and \ref{fig::lattice}(b).
For small doping level, the system is in the Luther-Emery liquid (LEL) phase~\cite{luther1974backward} except at $\delta = 1/4$, where a fully gapped CDW phase emerges due to the relevant umklapp interactions~\cite{balents1996,roux2007diamagnetism,white2002friedel}.
In the LEL phase, the system has a gapless charge mode and no gapless spin mode, which is also denoted as the C1S0 state.
Consequently, the SC pairing and charge density correlations decay algebraically, and the spin correlation and single-particle Green's function decay exponentially.
At $\delta \approx 0.4$, the system has a transition to the Tomonaga-Luttinger liquid (TLL) phase with both a gapless charge mode and a gapless spin mode, which is denoted as the C1S1 state with the corresponding correlation functions decaying algebraically~\cite{giamarchi2003quantum}.
At $\delta = 1/2$, the model is a unique insulating C0S1 state with a gapless spin mode~\cite{white2002friedel,lu2023ground}.

\subsection{Couple two LEL via $J_\bot$}
\label{subsec::LEL}

\subsubsection{Evolution to the interlayer SC with an intermediate CDW phase}

\begin{figure*}[htb]
\includegraphics[width=1.0 \linewidth]{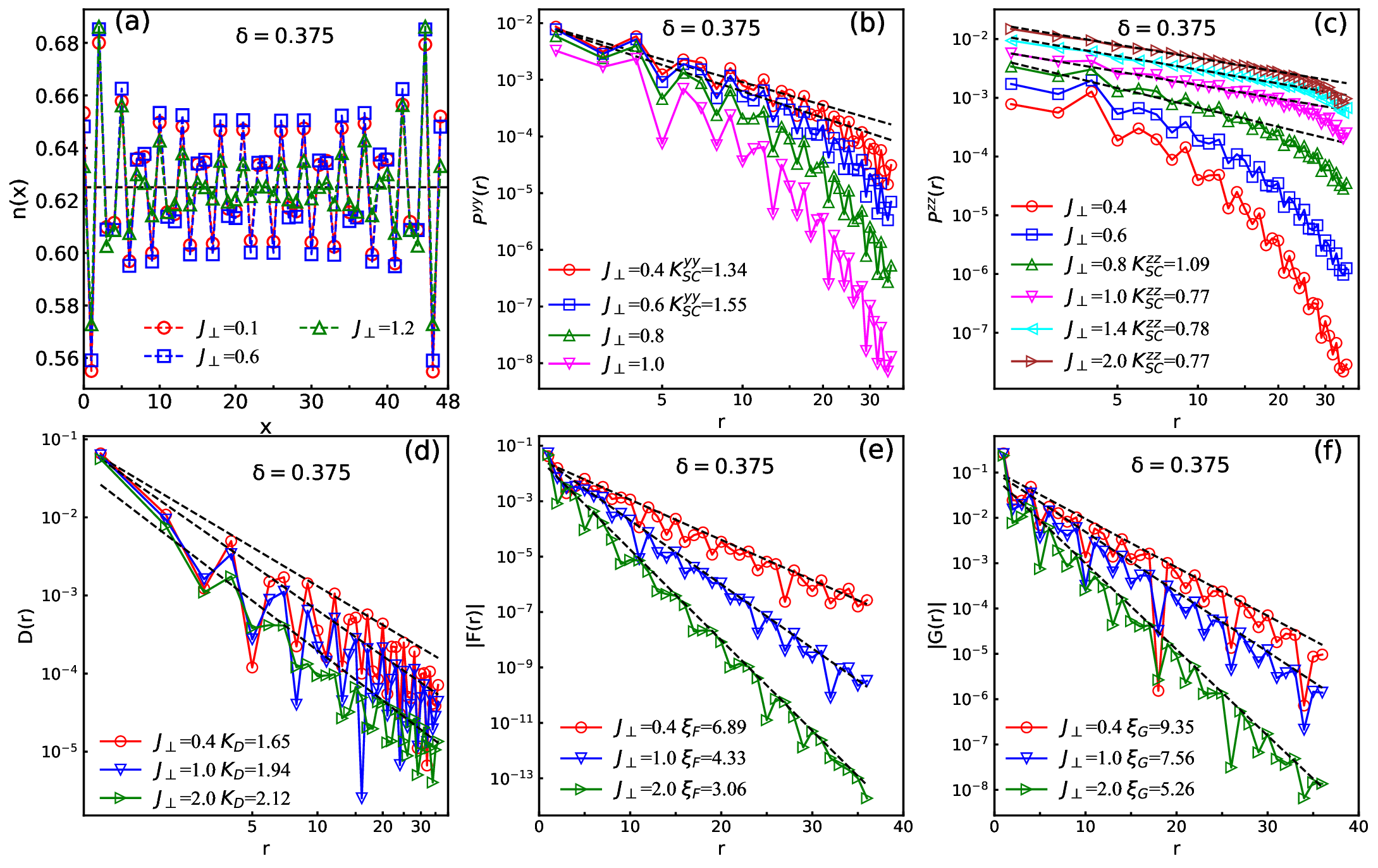}
\caption{Measurements of the bilayer two-leg $t$-$J$-$J_{\bot}$ model at $\delta=0.375$. (a) Local charge density $n(x)$. (b) In-plane pairing correlation $P^{yy}(r)$ (double-log scale). (c) Inter-plane pairing correlation $P^{zz}(r)$ (double-log scale). (d) Charge density-density correlation $D(r)$ (double-log scale). (e) Spin correlation $F(r)$ (semi-log scale). (f) Single-particle Green's function $G(r)$ (semi-log scale). The system length is $L_x=48$, and the number of kept SU(2) multiplets ranges from $6000$ to $12000$.}
\label{fig::0375_1}
\end{figure*}

We begin by studying the bilayer $t$-$J$-$J_{\bot}$ model at $\delta=0.1875$.
The two-leg $t$-$J$ ladder at this doping level is the LEL~\cite{lu2023ground}. 
In Fig.~\ref{fig::01875_1}, we present the DMRG results of the $t$-$J$-$J_{\bot}$ model with growing $J_{\bot}$.
The charge density profile $n(x)$ exhibits a strong CDW oscillation at $0.1 \leq J_{\bot} / J \lesssim 0.6$, but is strongly suppressed at larger $J_{\bot}$ [Fig.~\ref{fig::01875_1}(a)], which suggests a possible CDW order at small $J_{\bot}$.
In Figs.~\ref{fig::01875_1}(b) and \ref{fig::01875_1}(c), we show the pairing correlations $P^{yy}(r)$ and $P^{zz}(r)$. 
Clearly, the growing $J_{\bot}$ suppresses $P^{yy}(r)$ and enhances $P^{zz}(r)$.
While $P^{yy}(r)$ decays algebraically at $J_{\bot} = 0$ (in the LEL state), it appears to decay exponentially even at the small $J_{\bot} / J = 0.1$.
For $P^{zz}(r)$, it decays exponentially at small $J_{\bot}$, and the decay becomes algebraic at $J_{\bot} / J \gtrsim 0.8$ with very small power exponents $K_{{\rm{sc}}}^{zz} \sim 0.7$ indicating the divergent SC susceptibility at low temperature.
The emergence of the quasi-long-range inter-plane SC is associated with the drastic suppression of charge density oscillation.

Next, we analyze spin correlation and single-particle Green's function.
The spin correlation $F(r)$ always decays exponentially, characterizing the gapped spin mode.
In the inter-plane ZZ-SC phase at $J_{\bot} / J \gtrsim 0.8$, $F(r)$ is significantly suppressed [Fig.~\ref{fig::01875_1}(d)] because of the formed interlayer singlet pairing.
Similarly, $G(r)$ decays exponentially and is diminished with growing $J_{\bot}$ [Fig.~\ref{fig::01875_1}(e)].
To further characterize the different phases, we calculate the bipartite entanglement entropy. 
For a critical system described by the conformal field theory~\cite{calabrese2004entanglement,fagotti2011universal}, the entanglement entropy between the two subsystems with the lengths $x$ and $L_x - x$ follows the behavior
\begin{equation}
S(x) = \frac{c}{6}{\ln}[\frac{L_x}{\pi}\sin(\frac{x\pi}{L_x})] + g,
\label{eq::Sx}
\end{equation}
where $c$ is the central charge and $g$ is a non-universal constant. 
In Fig.~\ref{fig::01875_1}(f), the entropy data for $0.1 \leq J_{\bot} / J \lesssim 0.6$ exhibit a strong oscillation that is consistent with a stable CDW.
In this case, we follow the usual scheme in DMRG calculation by fitting the upper envelop of the bulk energy data~\cite{jiang2018superconductivity,jiang2019superconductivity,jiang2020topological,lu2023ground,peng2023charge}, giving a central charge $c \approx 0$ that agrees with a fully gapped CDW state emerging at commensurate fillings.
Notice that not only the upper envelop but also the whole entropy curve in the bulk of the system appears to be very flat.
On the other hand, for larger $J_{\bot}$ $c$ is very close to $1$, which is consistent with the C1S0 nature of the inter-plane ZZ-SC state.
The DMRG results identify an interlayer ZZ-SC phase at $J_{\bot} / J \gtrsim 0.8$ and strongly suggest an emergent CDW state at small $J_{\bot}$.

To further confirm this CDW state, we compare the results between the two-leg ladder (in the LEL state) and the weakly coupled case at $J_{\bot} / J = 0.1$.
For the LEL state on an open ladder, the charge density profile will exhibit the Friedel oscillation~\cite{white2002friedel,jiang2019superconductivity} described as
\begin{equation}
    n(x) = n_0 + A_{cdw}\cos(\frac{4\pi}{\lambda}x+\phi),
    \label{eq::nx_o}
\end{equation}
where $A_{cdw}=A_0[x^{-K_c/2}+(L_x+1-x)^{-K_c/2}]$ and $K_c$ characterizes the algebraic decay of the oscillation amplitude from the boundaries to the bulk.
The oscillation period $\lambda$ follows the LEL feature $\lambda = 1 / \delta$~\cite{lu2023ground}. 
We employ Eq.~\eqref{eq::nx_o} to fit the charge density profile and extract the power exponent $K_c$ [Fig.~\ref{fig::01875_2}(a)]. 
For the two-leg ladder, $K_c \sim 1.23$ agrees with the typical LEL behavior~\cite{lu2023ground}. 
In contrast, the bilayer model at $J_{\bot} / J = 0.1$ shows a robust CDW oscillation with a much smaller $K_c \sim 0.61$, which is close to the $K_c$ observed in the stripe phase of the single-layer $t$-$J$ model on the six-leg cylinder~\cite{gong2021robust}.
Furthermore, we zoom in and carefully compare the in-plane pairing correlation $P^{yy}(r)$ [Fig.~\ref{fig::01875_2}(b)].
While $P^{yy}(r)$ exhibits a good power-law decay in the two-leg ladder case, it drops much faster and follows an exponential decay at $J_{\bot} / J = 0.1$. 
In Fig.~\ref{fig::01875_1}(f), we have shown the entropy and the fitted central charge, which suggest a fully gapped state at small $J_{\bot}$.
In Fig.~\ref{fig::01875_2}(c), we further analyze the bond dimension dependence of entropy at $J_{\bot} / J = 0.1$ and also compare the results with the LEL of two-leg ladder. 
With growing bond dimension, the entropy data increase slightly, but the fitted central charges are always close to zero, which are quite different from the LEL state of the two-leg ladder. 
These carefully analyzed DMRG data consistently confirm fully gapped CDW states at small $J_{\bot}$ and commensurate fillings, between the in-plane SC (the LEL state) and inter-plane ZZ-SC state.
The two coupled LEL states are gapped at either infinitesimal $J_{\bot}$ or a very small $J_{\bot} / J < 0.1$, which is usually a Kosterlitz-Thouless phase transition~\cite{amit1980renormalisation}.
Given the numerical challenges of DMRG simulation at weak $J_{\bot}$ and identification of the Kosterlitz-Thouless transition, we will not address this issue in this work.

\subsubsection{Direct transition from in-plane to inter-plane SC phase}

When coupling two single-layer LEL states at $\delta=0.375$, the phase diagram with growing $J_{\bot}$ shows a notable difference from that at $\delta=0.1875$, as illustrated in Fig.~\ref{fig::lattice}(c). 
The system remains in the LEL phase up to $J_{\bot} / J \approx 0.6$. 
For the stronger $J_{\bot}$, the system has a transition to the ZZ-SC phase.
For $J_{\bot} / J \lesssim 0.6$, the oscillation amplitude of charge density profile $n(x)$ shows a visible decay from the open boundaries to the bulk [Fig.~\ref{fig::0375_1}(a)], which qualitatively agrees with the Friedel oscillation in the LEL.
For larger $J_{\bot}$, the oscillation amplitude of $n(x)$ is strongly suppressed, similar to the case at $\delta = 0.1875$.

Next, we compare the pairing correlations $P^{yy}(r)$ and $P^{zz}(r)$ [Figs.~\ref{fig::0375_1}(b) and \ref{fig::0375_1}(c)].
The decay of $P^{yy}(r)$ at small $J_{\bot}$ is close to the power-law behavior. 
For $J_{\bot} / J \gtrsim 0.8$, $P^{yy}(r)$ clearly becomes exponential decay.
Meanwhile, $P^{zz}(r)$ follows the exponential decay at $J_{\bot}  / J \lesssim 0.6$, but decays algebraically with small power exponents at $J_{\bot} / J \gtrsim 0.8$, suggesting a transition from the intralayer to interlayer SC.   
Furthermore, we present other correlation functions in Figs.~\ref{fig::0375_1}(d)-\ref{fig::0375_1}(f).
While density correlations $D(r)$ always appear like algebraic decay, spin correlation $F(r)$ and single-particle Green's function $G(r)$ keep the exponential decay, which are consistent with a transition from the in-plane to inter-plane SC.

\begin{figure*}[htb]
\includegraphics[width=1.0 \linewidth]{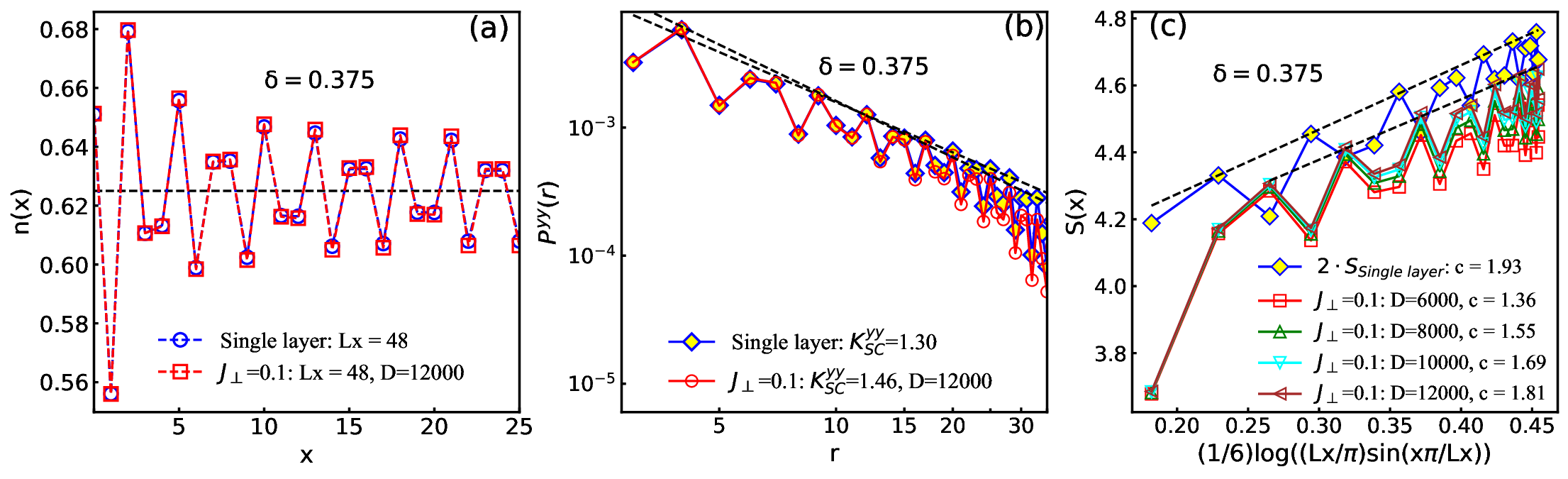}
\caption{Comparison of (a) local charge dnesity $n(x)$, (b) in-plane pairing correlation $P^{yy}(r)$, and (c) entanglement entropy $S(x)$ between the two-leg $t$-$J$ model and bilayer two-leg $t$-$J$-$J_{\bot}$ model at $\delta=0.375$. In (c), the blue line with yellow squares represents twice the entanglement entropy of the two-leg $t$-$J$ model, serving as a reference for the weakly coupled ($J_{\bot}/J=0.1$) bilayer system. Other curves show $S(x)$ of the $t$-$J$-$J_{\bot}$ model at different SU(2) multiplets bond dimensions. The central charge $c$ is extracted using Eq.~\eqref{eq::Sx}.
}
\label{fig::0375_2}
\end{figure*}

\begin{table}[t]
\centering
\setlength{\tabcolsep}{5pt}
\caption{The extracted power exponents $K_D$, $K_{\rm sc}^{yy}$ and $K_{\rm sc}^{zz}$ for different $J_{\perp}$ at $\delta=0.375$ in both the LEL and the ZZ-SC phase.}
\begin{tabular}{lccccc}
\toprule
Phase ($J_{\perp}/J$) & ~$K_D$ & ~$K_{\rm sc}^{yy}$ & ~$K_{\rm sc}^{zz}$  & ~$K_D K_{\rm sc}^{yy}$& ~$K_D K_{\rm sc}^{zz}$ \\
\hline
LEL (0.2)  & ~1.621 & ~1.248 & ~2.994  & ~2.023 & ~4.853 \\
LEL (0.4)  & ~1.652 & ~1.341 & ~2.405  & ~2.215 & ~3.973 \\
LEL (0.6)  & ~1.693 & ~1.546 & ~1.730  & ~2.617 & ~2.929 \\
ZZ-SC (1.0) & ~1.938 & ~2.507 & ~0.767 & ~4.859 & ~1.488 \\
ZZ-SC (1.4) & ~1.649 & ~5.158 & ~0.779 & ~8.506 & ~1.285 \\
ZZ-SC (1.8) & ~1.714 & ~5.670 & ~0.730 & ~9.718 & ~1.252 \\
\bottomrule
\end{tabular}
\label{Tab::1}
\end{table}

\begin{figure}[htb]
\includegraphics[width=0.8 \columnwidth]{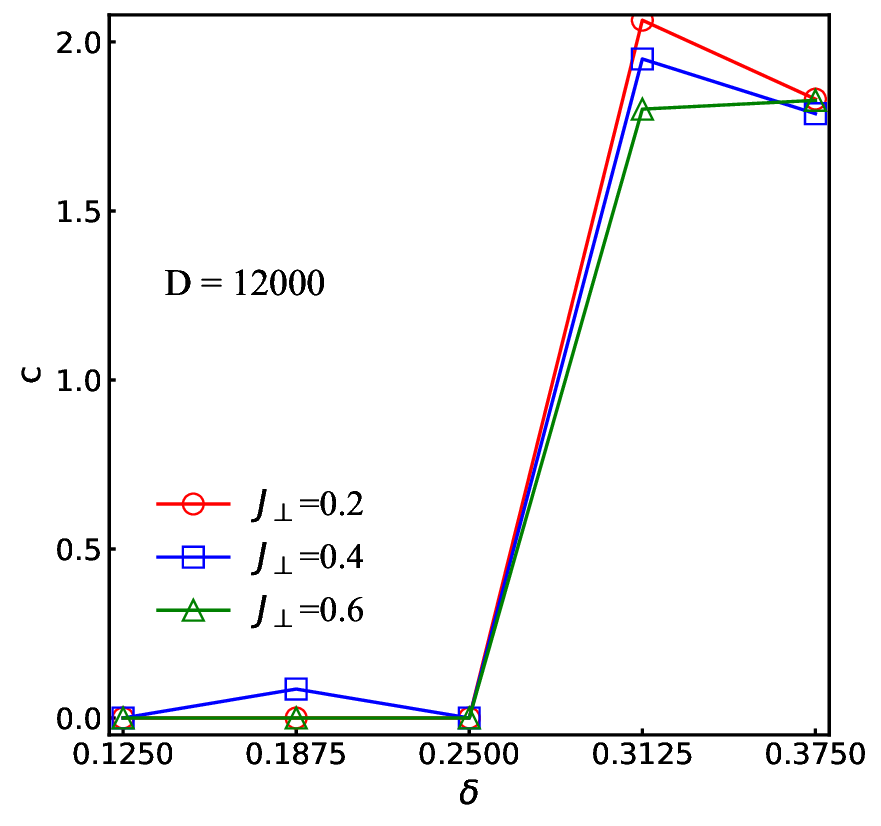}
\caption{Central charge extracted from the entanglement entropy $S(x)$ for the bilayer two-leg $t$-$J$-$J_{\bot}$ model in the weak coupling regime with $J_{\bot}/J = 0.2$, $0.4$ and $0.6$. The doping ratio $\delta$ varies from $0.125$ to $0.375$. The bond dimensions $D = 12000$. 
}
\label{fig::0375_3}
\end{figure}

While the two-leg $t$-$J$ ladder at $\delta = 0.375$ is the C1S0 LEL state, our results in Fig.~\ref{fig::0375_1} suggest that the in-plane C1S0 states remain stable for $J_{\bot} / J \lesssim 0.6$, which thus should give a C2S0 state for the bilayer system.
In Fig.~\ref{fig::0375_2}, we compare the results between the single-layer ladder and the $t$-$J$-$J_{\bot}$ model at $J_{\bot} / J = 0.1$, similar to Fig.~\ref{fig::01875_2}.
One can see that the charge density profiles $n(x)$ are almost the same for the two cases and $P^{yy}(r)$ are also very close, which are quite different from Fig.~\ref{fig::01875_2}.
By further examining the entanglement entropy with growing bond dimension, we find the fitted central charge at large bond dimensions approaches $c = 2$, supporting a C2S0 state at $J_{\bot} / J = 0.1$.
By comparing the results in Fig.~\ref{fig::01875_2} and Fig.~\ref{fig::0375_2}, we can distinguish the different states at small $J_{\bot}$ for $\delta = 0.1875$ and $\delta = 0.375$.

\begin{figure*}[htb]
\includegraphics[width=1.0 \linewidth]{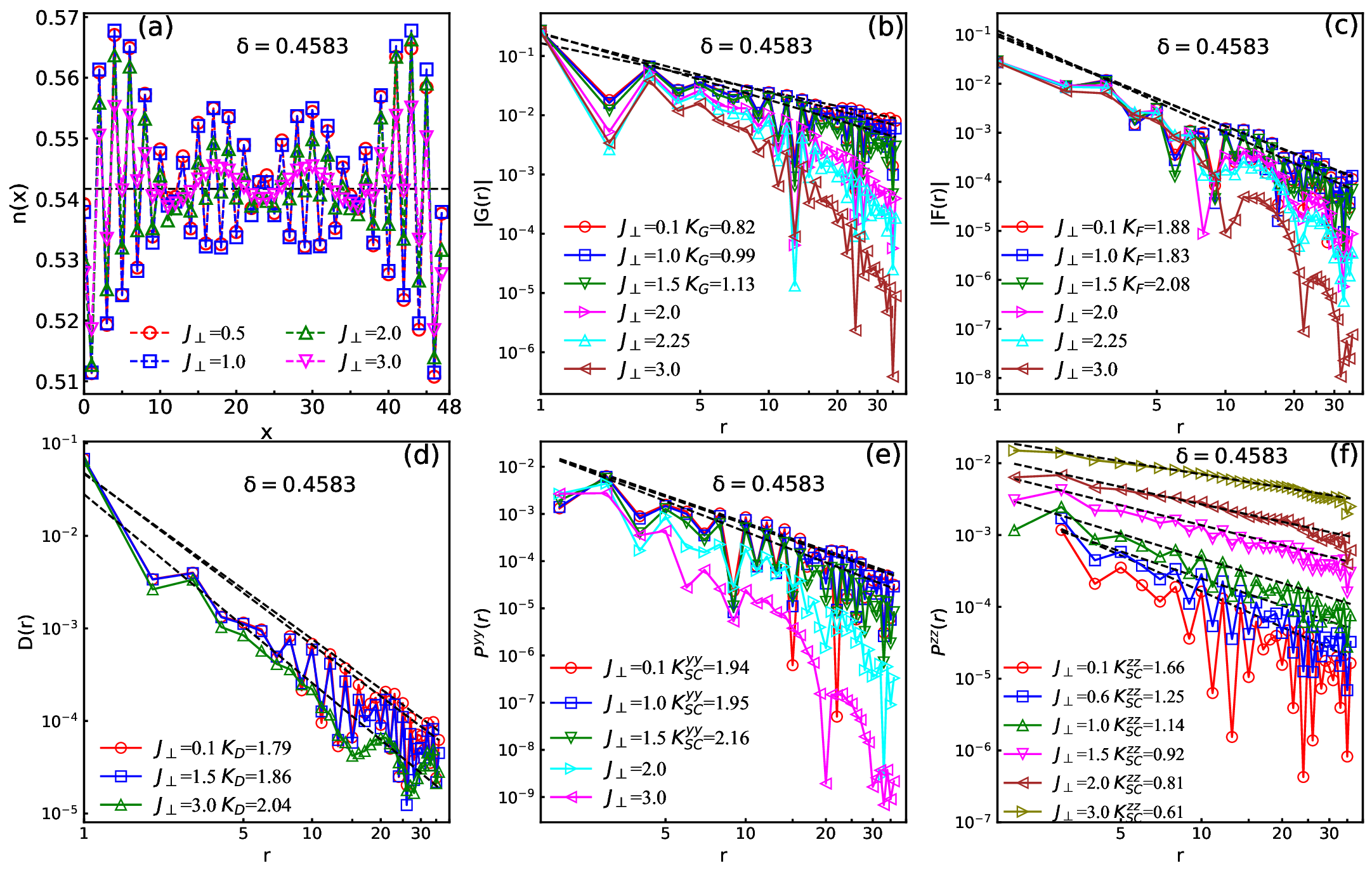}
\caption{Measurements of the bilayer two-leg $t$-$J$-$J_{\bot}$ model at $\delta=0.4583$. (a) Local charge density $n(x)$. (b) Single-particle Green's function $G(r)$ (double-log scale). (c) Spin correlation $F(r)$ (double-log scale). (d) Charge density correlation $D(r)$ (double-log scale). (e) In-plane pairing correlation $P^{yy}(r)$ (double-log scale). (f) Inter-plane pairing correlation $P^{zz}(r)$ (double-log scale). The system length is $L_x=48$, and the number of kept SU(2) multiplets ranges from $6000$ to $14000$.
}
\label{fig::04583_1}
\end{figure*}

To further clarify the nature of the LEL and ZZ-SC phases, we have extracted the power exponents of the density correlation function ($K_D$), the in-plane pairing correlation function ($K_{\rm sc}^{yy}$), and the interlayer pairing correlation function ($K_{\rm sc}^{zz}$) for several representative parameter points at $\delta = 0.375$ (see Table~\ref{Tab::1}).
In the LEL phase with in-plane SC, we find $K_D K_{\rm sc}^{yy} \gtrsim 2$ for $J_{\perp}/J \ge 0.2$.
Although the dual relation deviates from $K_D K_{\rm sc}^{yy} \sim 1$, this LEL phase seems to connect smoothly with the LEL phase of the single-layer system, as shown in Fig.~\ref{fig::0375_2}. 
Due to the suppressed in-plane SC with increasing $J_{\perp}$, the increased $K_{\rm sc}^{yy}$ also leads to the increase of $K_D K_{\rm sc}^{yy}$ while $K_D$ only grows slightly.
In contrast, in the ZZ-SC phase we find $K_D \lesssim 2$ and $K_{\rm sc}^{zz} \lesssim 1$, giving the product $K_D K_{\rm sc}^{zz}$ approaching $1$ with increasing $J_{\perp}$. 
This feature shows that the ZZ-SC phase at large $J_{\perp}$ can satisfy the dual relation of the LEL state.

\subsubsection{Mapping out the phase diagram}

Since the bilayer model at $\delta = 0.1875$ and $0.375$ hosts different ground states at small $J_{\bot}$, we further investigate the doping range of $\delta = 0.125 - 0.375$ with growing $J_{\bot}$.
As we have shown in Fig.~\ref{fig::01875_2}(c) and Fig.~\ref{fig::0375_2}(c), here the central charge could be an effective diagnosis of the different states, thus we fit the central charge at various doping levels.
The results by keeping $12000$ SU(2) bond dimensions at $J_{\bot} / J = 0.2,0.4,0.6$ are presented in Fig.~\ref{fig::0375_3}.
For $\delta = 0.125, 0.1875$, and $0.25$, the fitted central charge is close to zero, indicating a fully gapped state.
For $\delta = 0.3125$, the obtained central charge $c \sim 2$ is similar to $\delta = 0.375$.
Thus, we can identify the CDW states and the coupled LEL states (C2S0) as shown in Fig.~\ref{fig::lattice}(c).
Notice that in this phase diagram, we do not accurately determine the phase boundaries but only show the rough boundaries as the dashed lines.

For the two-leg $t$-$J$ ladder at most fillings, the system is either the LEL state (C1S0) or the Luttinger liquid state (C1S1) with a gapless charge mode~\cite{lu2023ground}.
At the commensurate filling $\delta = 0.25$, the system becomes a fully gapped CDW state characterized by the feature of one hole per CDW period in average~\cite{white2002friedel,lu2023ground}.
By checking the CDW pattern of the bilayer system, we find that this feature persists at our studied doping levels, indicating that these CDW states belong to the same state. 
The CDW state of the two-leg ladder at $\delta = 0.25$ extends to more doping levels when two layers are coupled. 
It should be noted that in such quasi-one-dimensional systems, the fully gapped CDW state may emerge only at specific commensurate fillings due to the relevant umklapp interactions~\cite{balents1996}. At incommensurate fillings when the CDW wavelength and lattice spacing mismatch, the CDW state has a gapless sliding mode~\cite{frohlich1954theory,ong1985generation}. Here, since we only consider the commensurate fillings, we obtain the fully gapped CDW state with $c \sim 0$.
Interestingly, at weak $J_{\perp}$ such as $J_{\perp}/J = 0.4$, the CDW phase gives way to a coupled LEL phase (C2S0) with increasing doping level, which may be understood as the system can gain energy by allowing the holes to pair and become mobile, rather than a static charge order in the higher doping range.

\subsection{Couple two TLL via $J_{\bot}$}
\label{subsec::TLL}

\begin{figure}[htb]
\includegraphics[width=1 \columnwidth]{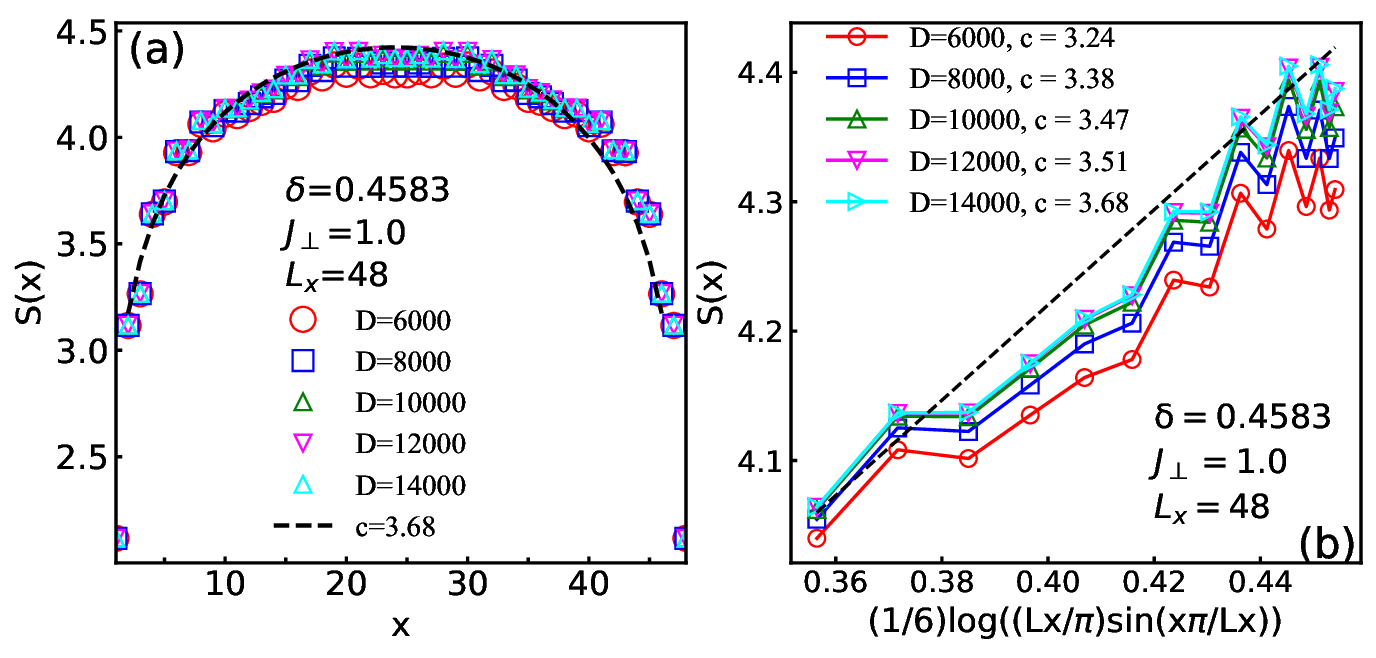}
\caption{Entanglement entropy $S(x)$ of the bilayer two-leg $t$-$J$-$J_{\bot}$ model at $\delta=0.4583$, $L_x = 48$, and $J_{\bot}/J = 1.0$. (a) and (b) show $S(x)$ versus the subsystem length $x$ and the conformal distance, respectively. $S(x)$ is fitted using Eq.~\eqref{eq::Sx} to extract the central charge $c$. The bond dimensions increase from $6000$ to $14000$. 
}
\label{fig::04583_2}
\end{figure}

\begin{figure*}[htb]
\includegraphics[width=1.0 \linewidth]{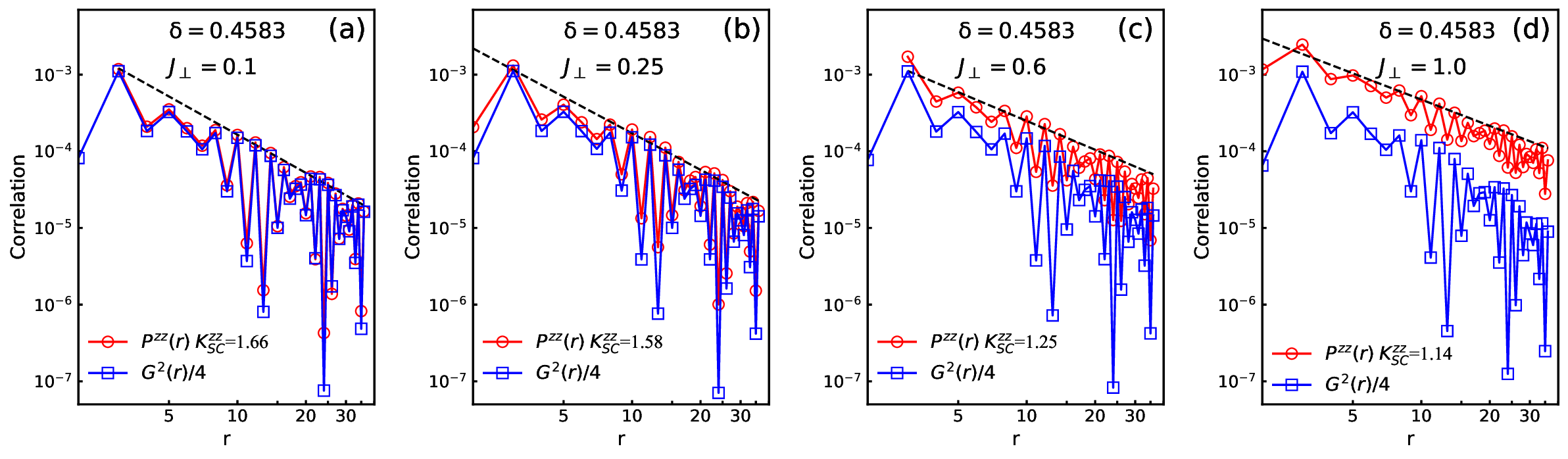}
\caption{Comparison of the inter-plane pairing correlation $P^{zz}(r)$ and the square of single-particle Green's function $G^2(r)/4$ at $\delta=0.4583$ with increasing $J_{\bot}$. (a) $J_{\bot}/J = 0.1$. (b) $J_{\bot}/J = 0.25$. (c) $J_{\bot}/J = 0.6$. (d) $J_{\bot}/J = 1.0$.
}
\label{fig::04583_3}
\end{figure*}

In recent studies, it has been shown that a strong $J_{\bot}$ can stabilize the ZZ-SC in the bilayer two- and three-leg $t$-$J$ ladders near $\delta = 0.5$~\cite{qu2024bilayer}.
In this subsection, we discuss the results by coupling two TLL states via $J_{\bot}$ from weak to strong interaction.
We choose $\delta = 0.4583$ ($\delta = 11/24$) as the representative, which for the two-leg ladder sits in the TLL phase (C1S1).

We first show the charge density profile $n(x)$ in Fig.~\ref{fig::04583_1}(a). 
With growing $J_{\bot}$, $n(x)$ is almost invariant up to $J_{\bot} / J \sim 1.5$, and for larger $J_{\bot}$ the amplitude of $n(x)$ is gradually suppressed, suggesting a possible phase transition at $J_{\bot} / J \sim 1.5$.
Below $J_{\bot} / J \sim 1.5$, both $G(r)$ and $F(r)$ exhibit power-law decay, with the power exponents $K_{\rm G} \sim 1$ and $K_{\rm F} \sim 2$, respectively [Figs.~\ref{fig::04583_1}(b) and \ref{fig::04583_1}(c)].
Above this $J_{\bot}$ value, $G(r)$ and $F(r)$ clearly become exponential decay.
Meanwhile, density correlation $D(r)$ always keeps the power-law decay with the power exponent $K_{\rm D} \sim 2$ [Fig.~\ref{fig::04583_1}(d)].

Furthermore, we compare $P^{yy}(r)$ and $P^{zz}(r)$ in Figs.~\ref{fig::04583_1}(e) and \ref{fig::04583_1}(f).
For $J_{\bot} / J \lesssim 1.5$, both $P^{yy}(r)$ and $P^{zz}(r)$ exhibit a good power-law decay, with the power exponents $K_{{\rm SC}}^{yy} \sim 2$ and $K_{{\rm SC}}^{zz} \lesssim 2$. 
Above this $J_{\bot}$ value, $P^{yy}(r)$ becomes exponential decay and $P^{zz}(r)$ shows a strong quasi-long-range order with the power exponent $K^{zz}_{\rm SC} < 1$.
The DMRG results unveil the stable in-plane Luttinger-liquid features below $J_{\bot} / J \lesssim 1.5$, and the ZZ-SC with gapped single-particle and spin excitations after the transition.
Furthermore, we calculate the entropy to fit the central charge.
Since the two-leg ladder is the C1S1 state with $c=2$, the coupled two C1S1 states should have $c = 4$ if no mode is gapped.
We demonstrate the entropy $S(x)$ at $J_{\bot} / J = 1.0$, $L_x = 48$ in Fig.~\ref{fig::04583_2}, with the data obtained from $6000$ to $14000$ SU(2) multiplets.
With approaching better convergence of entropy, the fitted central charge gets closer to $4$, which supports a C2S2 state of the bilayer model.

Interestingly, while the in-plane TLL features persist, the interlayer pairing correlation could also be enhanced.
As shown in Fig.~\ref{fig::04583_1}(f), the power exponent of $P^{zz}(r)$ decreases gradually with growing $J_{\bot}$ and shows a small value $K^{zz}_{\rm SC} \simeq 1.14$ at $J_{\bot} / J = 1.0$.
In Fig.~\ref{fig::04583_3}, we further compare $P^{zz}(r)$ and the single-particle correlation square $G^2(r)/4$.
For a weak interlayer coupling such as $J_{\bot} / J = 0.1$ and $0.25$, we find $P^{zz}(r) \approx G^2(r)/4$ indicating that the power-law decay of $P^{zz}(r)$ is solely from the contribution of $G(r)$.
With increasing $J_{\bot}$, $P^{zz}(r)$ is enhanced and becomes stronger than $G^2(r) / 4$ at the intermediate $J_{\bot}$, such as $K^{zz}_{\rm SC} \simeq 1.14$ at $J_{\bot} / J = 1.0$, which indicates the emergence of inter-plane SC while in-plane TLL features persist.

\section{Summary and discussion}
\label{sec::summary}

Motivated by the bilayer cuprate superconductors and the nickelate superconductor La$_3$Ni$_2$O$_7$, we have explored the ground-state properties of finite-width bilayer $t$-$J$-$J_{\bot}$ ladders, where $t$ is the in-plane electron hopping, $J$ is the in-plane spin  interaction, and $J_{\bot}$ is the interlayer spin exchange interaction. 
We map out a phase diagram of the bilayer two-leg ladders with tuning doping ratio $\delta = 1/8 - 1/2$ and $J_{\bot}$, by means of the density matrix renormalization group calculations.
The main purpose is to clarify how the in-plane SC correlations inherited from the single-layer Luther-Emery liquid or Luttinger liquid are transformed into inter-plane SC correlations with increasing $J_{\bot}$.
Notice that due to the DMRG convergence challenge of the bilayer three-leg ladders at small $J_{\bot}$, currently we cannot obtain an accurate phase diagram for the bilayer three-leg system and we only focus on the larger-$J_{\bot}$ regime without convergence issue.

When the single layer sits in either the LEL or TLL phase, a strong $J_{\bot}$ can always drive an inter-plane SC.
Interestingly, by coupling two LEL states, the evolution from in-plane to inter-plane SC depends on doping ratio.
At the smaller doping level, the LEL quickly becomes CDW states at either infinitesimal $J_{\bot}$ or a very small $J_{\bot}/J < 0.1$.
These CDW states seem to originate from the CDW state of the two-leg ladder at $\delta = 1/4$.
At the larger doping range, the LEL remains stable up to the intermediate $J_{\bot}$ and then has a direct transition to the inter-plane SC phase.
In this direct evolution, our DMRG results do not find a coexistence regime of in-plane and inter-plane SC.
For the two-leg ladder in the Luttinger liquid phase, a larger $J_{\bot} / J$ is required to drive the interlayer SC phase. 
Interestingly, at the intermediate $J_{\bot}/J$ regime (before the transition to the ZZ-SC phase) such as $J_{\bot} / J = 1.0$, the pairing correlation $P^{zz}(r)$ is much stronger than the single-particle correlation square and $P^{zz}(r)$ shows a quasi-long-range order with the power exponent $K^{zz}_{\rm SC} \sim 1$, which strongly suggests the emergence of inter-plane SC while the in-plane Luttinger-liquid features persist.

At $\delta = 0.5$, the absence of SC originates from the C0S1 nature of the two-leg $t$-$J$ ladder, which is equivalent to an effective spin-$1/2$ chain~\cite{lu2023ground}.
Thus, couple two such two-leg ladders by a strong $J_{\perp}$ will gap the spin mode and lead to a fully gapped state~\cite{qu2024bilayer}.
Therefore, the absence of SC at $\delta = 0.5$ in the bilayer two-leg system is a special consequence of coupling two C0S1 states, rather than a generic feature of wider bilayer ladders.
For the wider case with $L_y = 3$, the previous DMRG studies have shown an interlayer SC near $\delta = 0.5$ (including $\delta = 0.5$), driven by a strong $J_{\bot}$ coupling~\cite{qu2024bilayer}.
We further examine the smaller doping level at $\delta = 0.375$, in which the three-leg ladder sits in the LEL phase, and the inter-plane SC at strong $J_{\bot}$ is still robust [see Appendix~\ref{appen:three}], showing the stable inter-plane SC for both $L_y = 2$ and $L_y = 3$ over a wide range of doping ratio, as long as $J_{\bot}$ is strong.
To confirm the ZZ-SC state of $\delta = 0.5$ at large $J_{\bot}$ (beyond $L_y = 2$), we also examine one point ($J_{\bot} = 3.0$) of the bilayer four-leg ladders ($L_y = 4$) [see Fig.~\ref{fig::Ly4}], which shows the same ZZ-SC behavior  demonstrating the robust inter-plane SC in wider systems.

In our phase diagram, we have considered a relatively large $J_{\perp}$.
Notice that in the single-layer $t$-$J$ model, a strong spin-exchange coupling $J$ can induce a phase separation~\cite{Rommer2000}. 
To examine the possibility of phase separation in our phase diagram, we have analyzed the charge density distributions for all the simulated parameter points, which show no evidence of formed charge clusters even for a large $J_{\perp}$, characterizing the absence of phase separation in the studied parameter region.
Compared with the single-layer $t$-$J$ model, the absent phase separation in this system may be owing to the small in-plane spin-exchange coupling ($t/J = 3$) and the bilayer structure with two equivalent layers.
In other multi-layer systems such as the trilayer $t$-$J$-$J_{\perp}$ model, the strong $J_{\perp}$ may drive a phase separation among the different layers.

At last, we discuss the potential connection between our results and related superconductors. 
For the nickelate La$_3$Ni$_2$O$_7$ under high pressure, DFT calculations suggested a strong interlayer coupling~\cite{luo2023bilayer}, which in this $t$-$J$-$J_{\perp}$ framework could be consistent with the ZZ-SC phase near $\delta = 1/2$, with the dominant interlayer $d_{x^2-y^2}$ orbital pairing.
For bilayer cuprates, the interlayer exchange $J_{\perp}$ is usually much weaker than the intralayer coupling $J$~\cite{tranquada1989neutron,haug2010neutron}. 
For this case near $1/8$ doping, the bilayer two-leg $t$-$J$ model exhibits a CDW phase without SC.
We have also examined the bilayer three-leg systems at weak $J_{\perp}$, which appear to be more challenging to converge and we do not deeply explore this regime in this work. 
For future study, it would be important to check weather the in-plane $d$-wave SC can be stabilized in the small $J_{\perp}$ regime of the wider bilayer systems.  

\begin{acknowledgments}
We acknowledge the stimulating discussion with Rong-Yang Sun.
Y.~Y. was supported by the China Postdoctoral Science Foundation under Grant No. 2024M763126. X.~L. and S.~S.~G. were supported by the National Science Foundation of China (No. 12534009 and No. 12274014), the Guangdong Provincial Quantum Science Strategic Initiative (No. GDZX2501006), the Special Project in Key Areas for Universities in Guangdong Province (No. 2023ZDZX3054), and the Dongguan Key Laboratory of Artificial Intelligence Design for Advanced Materials.
Y. W. was supported by the National Natural Science Foundation of China (NSFC, Grants No. 12547112 and No. 12188101) and by the Chinese Academy of Sciences through the Project for Young Scientists in Basic Research (Grant No. YSBR-059).
W. Q. C. was supported by the National Key Research and Development Program of China (No. 2024YFA1408101), NSFC (Grants No. 12141402, 12334002), Guangdong Provincial Quantum Science Strategic Initiative Grand No. SZZX2401001, the SUSTech-NUS Joint Research Program, the Science, Technology and Innovation Commission of Shenzhen Municipality (No. ZDSYS20190902092905285), and Center for Computational Science and Engineering at Southern University of Science and Technology.
The computational resources were also supported by the SongShan Lake HPC Center (SSL-HPC) in Great Bay University.
\end{acknowledgments}

\appendix

\section{Couple two C0S1 states via $J_{\bot}$}
\label{subsec::C0S1}

\begin{figure*}[htb]
     \includegraphics[width=1.0 \linewidth]{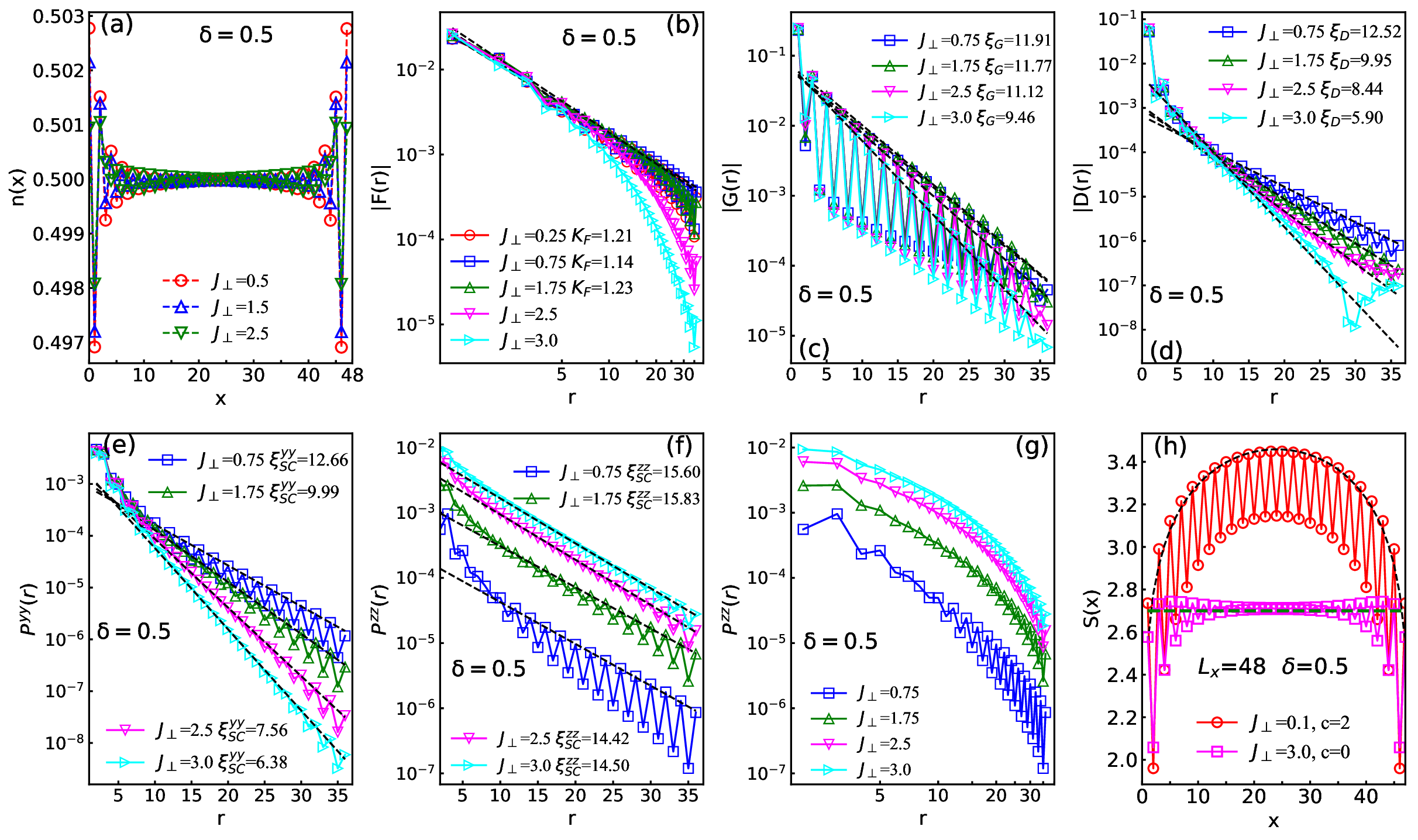}
\caption{Measurements of the bilayer two-leg $t$-$J$-$J_{\bot}$ model at $\delta=0.5$. (a) Local charge density $n(x)$. (b) Spin-spin correlation $F(r)$ (double-log scale). (c) Single-particle Green's function $G(r)$ (semi-log scale). (d) Charge density correlation $D(r)$ (semi-log scale). (e) Inter-plane pairing correlation $P^{yy}(r)$ (semi-log scale). (f) and (g) show the inter-plane pairing correlation $P^{zz}(r)$ in semi-log and double-log scales, respectively. (h) Entanglement entropy $S(x)$. The system length is $L_x=48$, and the number of kept SU(2) multiplets ranges from $6000$ to $12000$.
}
\label{fig::C0S1}
\end{figure*}

\begin{figure*}[htb]
\includegraphics[width=1.4 \columnwidth]{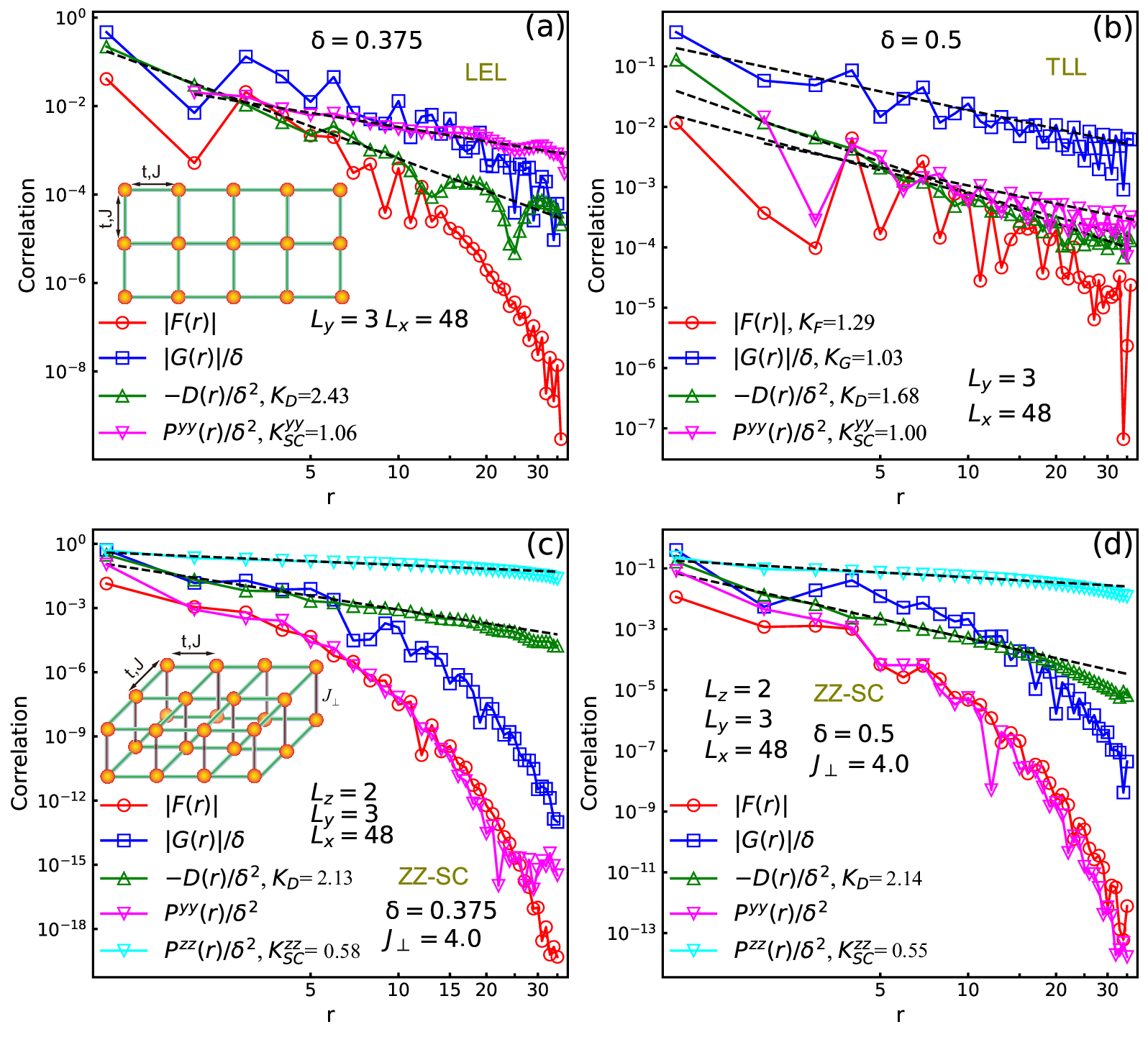}
\caption{Correlation functions of the three-leg $t$-$J$ model and bilayer three-leg $t$-$J$-$J_{\bot}$ model. (a) Double-log plots for the three-leg system at $\delta=0.375$. (b) Double-log plots for the three-leg system at $\delta=0.5$. (c) Double-log plots for the bilayer three-leg $t$-$J$-$J_{\bot}$ model at $\delta=0.375$ and $J_{\bot}/J = 4.0$. (d) Double-log plots for the bilayer three-leg $t$-$J$-$J_{\bot}$ model at $\delta=0.5$ and $J_{\bot}/J = 4.0$. The system length is $L_x=48$, and the number of kept SU(2) multiplets ranges from $6000$ to $8000$. 
} 
\label{fig::Ly3}
\end{figure*}

At $\delta = 0.5$, the single-layer two-leg $t$-$J$ ladder is the special C0S1 state with a gapless spin mode and gapped charge mode [Fig.~\ref{fig::lattice}(b)]. 
This state can be considered as an effective spin-$1/2$ chain. 
By employing the $J_{\bot}$ coupling between the two-leg ladders, one can expect the fully gapped spin mode above a critical $J_{\bot}$.
Since SC will not emerge, this case was not systematically studied in previous DMRG calculations.
Here, in order to compare with the case by coupling two three-leg ladders, we demonstrate the DMRG results for $\delta = 0.5$ in Fig.~\ref{fig::C0S1}.

While charge density remains $0.5$ with growing $J_{\bot}$, spin correlation changes the decay behavior from power-law to exponential at $J_{\bot} / J \sim 2$, showing the gapped spin mode.
Meanwhile, single-particle Green's function, density and pairing correlations always decay exponentially.
In particular, the interlayer pairing correlation $P^{zz}(r)$ remains exponential decay at large $J_{\bot}$ although its magnitude increases with growing $J_{\bot}$ [see Figs.~\ref{fig::C0S1}(f) and \ref{fig::C0S1}(g)].  
This phase transition from two coupled C0S1 states to a fully gapped state can also be characterized by the central charge, which changes from $c = 2$ to $c = 0$, as shown in Fig.~\ref{fig::C0S1}(h) for two typical points.
Thus, the absence of ZZ-SC at $\delta = 0.5$ in the bilayer two-leg system can be understood from the coupling of two C0S1 states.

\section{Couple two three-leg ladders via $J_{\bot}$}
\label{appen:three}

While $\delta = 0.5$ is a special doping ratio for two-leg ladder, ZZ-SC has been shown to exist when strongly coupling two three-leg ladders via $J_{\bot}$ at $\delta = 0.5$~\cite{qu2024bilayer}.
Here, we compare two different doping levels $\delta = 0.375$ and $0.5$ in the case of coupling two three-leg ladders, i.e. $L_y = 3$.
First, we examine the ground state properties of the single-layer three-leg $t$-$J$ model. 
As shown in Fig.~\ref{fig::Ly3}(a) for $\delta=0.375$, both $F(r)$ and $G(r)$ exhibit the exponential decay, but $D(r)$ and $P^{yy}(r)$ follow the power-law decay, which characterize the system as a LEL state, similar to the two-leg $t$-$J$ ladder at $\delta = 0.375$. 
For $\delta=0.5$, the three-leg $t$-$J$ model exhibits the TLL features with algebraic correlation functions [Fig.~\ref{fig::Ly3}(b)].

When two three-leg $t$-$J$ ladders are coupled, we find the DMRG simulations at weak $J_{\bot}$ are hard to converge, thus here we only focus on the strong $J_{\bot}$ regime.
For both $\delta = 0.375$ and $\delta = 0.5$, the ZZ-SC phase can emerge at strong $J_{\bot}$, as shown in Figs.~\ref{fig::Ly3}(c) and \ref{fig::Ly3}(d) at $J_{\bot} / J = 4$. 
While the density correlation $D(r)$ and interlayer pairing correlation $P^{zz}(r)$ exhibit power-law decay, $F(r)$, $G(r)$ and $P^{yy}(r)$ decay exponentially, showing the stable ZZ-SC by coupling two three-leg $t$-$J$ ladders via a strong $J_{\bot}$.
In particular, the bilayer three-leg result at $\delta = 0.5$ demonstrates that the ZZ-SC state at strong $J_{\bot}$ is recovered once the special $L_y = 2$ case is avoided.

\begin{figure}[htb]
\includegraphics[width=1.0 \columnwidth]{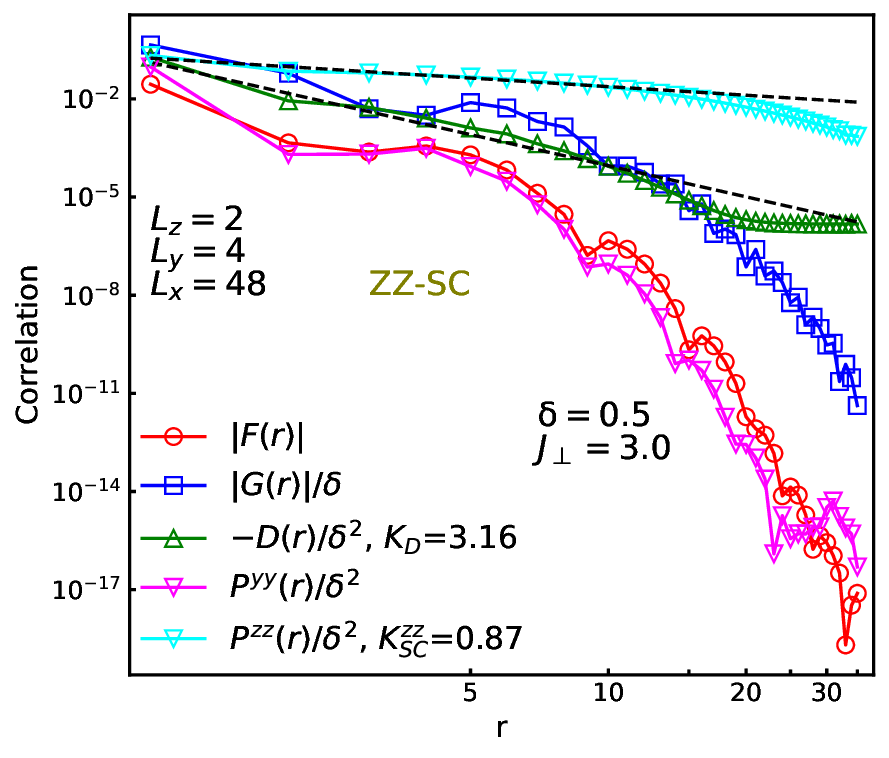}
\caption{Correlation functions of the bilayer four-leg $t$-$J$-$J_{\bot}$ ladders with $\delta = 0.5$, $J_{\bot}/J = 3.0$, and $L_x=48$. The number of kept SU(2) multiplets is $10000$, and the truncation error is about $5 \times 10^{-5}$.
}
\label{fig::Ly4}
\end{figure}

\section{Couple two four-leg ladders via a strong $J_{\bot}$ at $\delta = 0.5$}
\label{appen:four}

To further clarify that the absent ZZ-SC of the strongly coupled bilayer two-leg ladders at $\delta = 0.5$ is only a special case, we show the results of the bilayer four-leg ladders at $\delta = 0.5$ and a strong $J_{\bot}$ ($J_{\bot}/J = 3.0$).
The bilayer four-leg ladders have eight sites in each column, which is very hard to converge in DMRG calculation for a general case. 
Here, the potential ZZ-SC state at a large $J_{\bot}$ is relatively easier to converge as some modes are fully gapped.
Therefore, we examine the system ($J_{\bot}/J = 3.0$) by keeping $10000$ SU(2) multiplets, giving the truncation error about $5 \times 10^{-5}$.

The obtained correlation functions are shown in Fig.~\ref{fig::Ly4}, which are all consistent with the results at $L_y = 3$. The density correlation $D(r)$ and inter-plane pairing correlation $P^{zz}(r)$ exhibit power-law decay, while the spin correlation $F(r)$, single-particle Green's function $G(r)$, and intra-layer pairing correlation $P^{yy}(r)$ decay exponentially. These results indicate that a strong $J_{\bot}$ can also stabilize a quasi-long-range ZZ-SC phase in the bilayer four-leg ladders, supporting a robust ZZ-SC at $\delta = 0.5$ in wider systems.

%

\end{document}